\documentclass[prl,showpacs,10pt,twocolumn,superscriptaddress]{revtex4-1}
\usepackage{braket}
\usepackage{amsmath}
\usepackage{color}
\usepackage{graphicx}% Include figure files
\usepackage{dcolumn}% Align table columns on decimal point
\usepackage{bm}% bold math
\begin{document}

\title{The intermediate and spin-liquid phase of the half-filled honeycomb Hubbard 
model}
%Theory Study of Hubbard Model on Honeycomb Lattice}
\author{Qiaoni Chen}
\affiliation{Department of Chemistry, Princeton University, Princeton, NJ 08544}
\author{George~H.~Booth}
\affiliation{Department of Chemistry, Princeton University, Princeton, NJ 08544}
\author{Sandeep Sharma}
\affiliation{Department of Chemistry, Princeton University, Princeton, NJ 08544}
\author{Gerald Knizia}
\affiliation{Department of Chemistry, Princeton University, Princeton, NJ 08544}
\affiliation{Institut fuer Theoretische Chemie, Universitaet Stuttgart, Pfaffenwaldring 55, D-70569 Stuttgart, Germany}
\author{Garnet Kin-Lic Chan}
\affiliation{Department of Chemistry, Princeton University, Princeton, NJ 08544}
\affiliation{Department of Physics, Princeton University, Princeton, NJ 08544}

\begin{abstract}
We obtain the phase-diagram of the half-filled honeycomb Hubbard model with density matrix embedding theory, 
to address recent controversy at intermediate couplings. We use clusters from 2-12 sites and lattices at the thermodynamic limit.
We identify a paramagnetic insulating state, with possible hexagonal cluster order, competitive with the antiferromagnetic
phase at intermediate coupling. However, its stability is strongly cluster and lattice size dependent,
explaining controversies in earlier work. Our results support the paramagnetic insulator as being a metastable, rather than a true, intermediate phase,
in the thermodynamic limit.
\end{abstract}

\pacs{71.10.-w, 71.10.Fd, 71.30.+h, 75.10.Kt, 71.20.-b, 71.15.-m}

\maketitle

Recently, there has been much debate over the zero-temperature phases of the 
Hubbard honeycomb model at half-filling. The accepted picture for many years 
was that for small interactions $U$, the system is semi-metallic (SM) with a 
Dirac cone in the density of states~\cite{PhysRevB.87.205127}; at large $U$, the system is antiferromagnetically long-range ordered (AFM)~\cite{Sorella1992_EPL19-699,Martelo1997_ZPB103-335--338}. 
However, Meng {\it et al.}  proposed recently that an {\it additional}  phase appears at intermediate couplings, arising from strong quantum 
fluctuations due to the low coordination number~\cite{Meng2010_Nature464-847--851}. This phase has been further suggested to be a {\it gapped spin liquid}~\cite{Wang2010_PRB82-024419,Xu2011_PRB83-024408,Lu2011_PRB84-024420,Clark2013_ArXive-prints1305-0278}. 
If present, this would be extremely significant, as spin liquids are not known to exist at half-filling without
frustration~\cite{Nature.Balents.2010,Kimchi2013_PNAS110-16378-16383,PhysRevLett.110.096402}, and would greatly advance the search for experimental realizations 
of 2D spin-liquids,
both in correlated honeycomb materials as well as optical lattices~\cite{Novoselov2004_Science306-666--669,PhysRevA.87.043613}.

A large number of numerical methods have been applied to study the honeycomb Hubbard model~\cite{Sorella1992_EPL19-699,Martelo1997_ZPB103-335--338,Drut2009_PRL102-026802,Drut2009_PRB79-165425,
Paiva2005_PRB72-085123,Lee2009_PRB80-245118,Meng2010_Nature464-847--851,Clark2011_PRL107-087204,Yang2012_NJP14-115027,Sorella2012_SciRep2-992,Tran2009_PRB79-125125,Jafari2009_EPJB68-537-542,Wu2010_PRB82-245102,Liebsch2011_PRB83-035113,Seki2012_ArXive-prints1209-2101,
He2012_PRB86-045105,PhysRevB.87.205127,PhysRevLett.110.096402,PhysRevLett.107.010401,PhysRevB.85.205102,Li2011_EPL93-37007,Giuliani2010_CMP293-301-346,Assaad2013_PRX3-031010}. 
The work by Meng {\em et al.} used zero-temperature auxiliary-field (determinant) quantum Monte Carlo (AFQMC)~\cite{Meng2010_Nature464-847--851}. 
These results were viewed with particular confidence because AFQMC has no sign problem in this model 
and thus correlations were treated ``exactly'',  the only errors arising from using a finite size lattice of 648 sites.
Subsequent to this, many reports of an intermediate phase have appeared using quantum cluster methods~\cite{RevModPhys.77.1027}, such as Cluster 
Dynamical Mean-Field Theory (CDMFT) with exact diagonalization (ED) (Liebsch~\cite{Liebsch2011_PRB83-035113} and He {\it et al.}~\cite{He2012_PRB86-045105}) and quantum Monte Carlo (CT-QMC) solvers (Wu {\it et al.})~\cite{PhysRevB.85.205102}, and the Variational Cluster Approximation (VCA)~\cite{PhysRevLett.107.010401}. While these various quantum cluster methods differ in some details, they share a 
unifying Green function and self-energy functional formulation~\cite{Potthoff2003_EPJB32-429--436,Potthoff2003_EPJB36-335--348}, thus we refer to them collectively as 
Green function cluster (GFC) methods.

Despite these initial reports, there now appears increasing evidence that there may in fact be no spin-liquid phase~\cite{Sorella2012_SciRep2-992,PhysRevLett.110.096402,Clark2011_PRL107-087204}! The strongest hint is from 
Sorella {\it et al.}~\cite{Sorella2012_SciRep2-992}, who repeated Meng's AFQMC calculation with a larger lattice of 2592 sites.
%% Sorella {\it et al.}  also obtained the spin gap directly rather than the inferring it
%% from  long-range decay of the spin-correlation functions used by Meng {\it et al}.  
They found that at the increased lattice size, the region for an
intermediate phase shrank significantly, suggesting weak (if any) evidence for a spin-liquid phase. Subsequently, a further VCA calculation by Hassan and Senechal, in contrast to
earlier quantum cluster calculations, also found no spin-liquid phase~\cite{PhysRevLett.110.096402}.

These conflicting reports raise important questions both for the physics of the honeycomb Hubbard model, and
the numerical methods used to study it. Is there an intermediate phase, and is it a spin-liquid? If not, what is  observed
in calculations which  see an intermediate phase - is the system ``close'' to a spin-liquid or some
other  state? Why do some quantum cluster calculations observe an intermediate phase and others not? And how do the 
various numerical approximations, such as  finite lattice size in AFQMC, or finite cluster size in GFC methods, bias the calculations?
These are the main questions we target in this work.

To answer these questions, we use a different quantum cluster method to those used previously - the recently proposed Density Matrix Embedding Theory (DMET)~\cite{Knizia2012_PRL109-186404,Knizia2013_JCTC9-1428-1432,Booth2013_ArXive-prints1309-2320,Scuseria_Arxiv}. Note that DMET is not a variant of the GFC methods  discussed above.
%% , which provides an independent perspective on existing GFQC calculations. 
In principle, GFC methods complement AFQMC, because they work in the thermodynamic limit, but only
treat a finite range of correlations determined by the impurity cluster size. In practice, however, solving the
impurity problem within GFC methods involves further numerical approximations, such as bath 
discretization, or analytic continuation. Both require careful treatment to avoid affecting the physics. 
%% For example, Hassan {\it et al.}~\cite{PhysRevLett.110.096402} showed that poor bath discretization leads to an erroneous elimination of the semi-metal phase at small $U$.
In contrast, DMET is a quantum cluster method without bath discretization error by construction, and which yields a quantum impurity problem where spectral functions can be 
practically computed without analytic continuation. This allows us to study the honeycomb Hubbard model as a function of cluster correlation length and at
the thermodynamic limit with no further extraneous numerical approximations.

%% For example, if ED is used as a solver, as in Refs.~\cite{honeycombed}, 
%% one must discretize the bath degrees of freedom. The
%% importance of the bath discretization error was highlighted recently by Hassan {\it et al.}~\cite{hassan}, who critically examined VCA and CDMFT calculations 
%% with a single bath orbital per impurity site. These calculations can fail to even reproduce a semi-metal (SM) phase, producing a gap for all values of the 
%% interaction strength\cite{PhysRevLett.110.096402,Seki2012_ArXive-prints1209-2101}. If QMC is used as a solver, bath discretization error is avoided, but
%% the single-particle gap  used to identify the spin-liquid phase must be obtained by analytic continuation and extrapolation to zero-temperature, 
%% incurring additional errors. 
%% In total, such numerical issues potentially confuse resolving
%% whether different phases in the GFQC calculations appear due to numerical artifacts from solving the quantum impurity problem, or due to 
%% bias from the short-range correlations of the cluster approximations.

DMET maps a large lattice problem to a quantum impurity plus bath problem, and is exact in the weak- and strong-coupling limits, as well as the limit
of infinite cluster size.
The main conceptual difference between DMET and GFC methods~\cite{PhysRevLett.62.324,PhysRevLett.69.1240,RevModPhys.78.865,RevModPhys.68.13}, is that the 
DMET bath is derived from the {\it entanglement} of the  quantum state of the 
 impurity cluster, rather than its single-particle Green function. 
DMET has several important technical advantages. If one is interested in
thermodynamic properties, no Green functions need  be computed;  only
 the ground-state quantum impurity problem is solved, a significant computational savings.
Further, since the DMET bath is defined from a Schmidt decomposition, the bath is {\it the same size as the
impurity with no bath discretization error}, in contrast to the infinite bath in DMFT.
%% Because the DMET quantum impurity problem is so small and simple, one can solve it exactly without additional approximations.
%% For spectral properties, as obtained with spectral DMET, the Schmidt decomposition of the response vector shows that the  bath to 
%% reproduce the dynamical entanglement at each frequency is also of the same size as the 
%% number of impurity sites. The key point, however, is that there is a different bath for each frequency,
%% and this allows a continuous spectrum to be produced by a finite set of bath sites without unphysical broadenings. 
%% Further, the small size of the frequency-dependent DMET quantum impurity problem means that exact diagonalization can be used to obtain the spectral functions on the real axis
%% without analytic continuation.
Typically, an $n$-site cluster ($n>1$) DMET calculations yields similar physics to an $n$-site CDMFT that is converged with
respect to the bath parametrization, even though the DMET calculations require only a fraction of the cost~\cite{Knizia2012_PRL109-186404,Knizia2013_JCTC9-1428-1432,Booth2013_ArXive-prints1309-2320,Scuseria_Arxiv}.

\begin{figure}[!t]
\includegraphics[width=8.6cm]{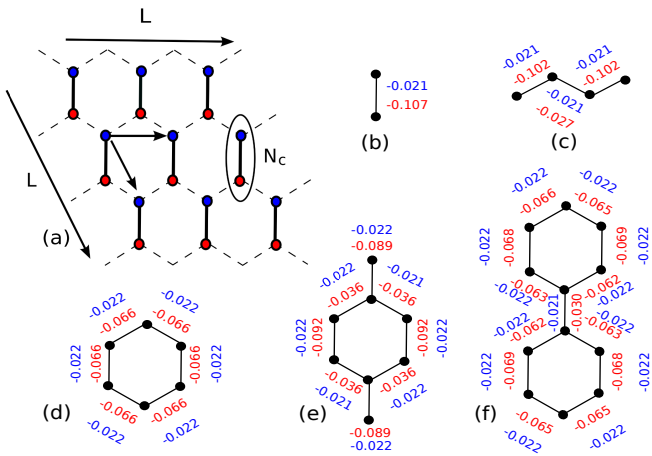}
\caption{(Color online) (a) DMET embeds clusters of $N_c$ sites in the underlying
honeycomb lattice of $2L^2$ sites. (b)-(f) Cluster shapes in our study, for $N_c= 2-12$. Also shown is the nearest-neighbour spin-spin correlation, $\langle S^z_i S^z_j \rangle$. Blue numbers are from $U=1.5t$ in the semi-metal (SM) phase,  red numbers  from $U=8.4t$ in the paramagnetic insulator (PMI) phase.}
\label{fig:cluster}
\end{figure}

We now describe the DMET method briefly; for more details we refer to the original references~\cite{Knizia2012_PRL109-186404,Knizia2013_JCTC9-1428-1432,Booth2013_ArXive-prints1309-2320}.
We first choose an impurity cluster, cut from the underlying lattice.
The underlying lattice is finite, but can trivially be made very large: in this study, we use lattices with more than 90,000 sites.
%% The ability to study lattices of different size, however, allows us to use
%% DMET to understand the finite size errors observed in other calculations, for example in AFQMC.
The ``environment'' lattice sites outside the cluster are then replaced by a bath. In ground-state DMET, the bath is defined
with the help of a model lattice ground-state wavefunction  $\ket{\Psi^{(0)}}$.
In this study, $\ket{\Psi^{(0)}}$ is a (possibly spin-broken) Slater determinant (ground-state) of a non-interacting lattice Hamiltonian $h + u$, 
where $h$ is the hopping matrix, and $u$ is a frequency-independent one-particle operator acting in each cluster cell on the lattice, analogous to a cluster self-energy. 
For ground-state DMET, the Schmidt decomposition of $\ket{\Psi^{(0)}}$ between the cluster and environment defines a bath space
$\{ |\beta^0\rangle\}$, of the same size as the impurity space.
%%  between
%% the  impurity cluster and its environment defines a set of impurity states $\{ \ket{\alpha_i}\}$
%% {\it and the same number of bath states} 
The quantum impurity Hamiltonian $H'$ is then obtained by projecting
a model lattice Hamiltonian $H_{lat}$ onto the quantum impurity plus bath space,
\begin{align}
H'&=P H_{lat} P \notag\\
&=P (h + U_{cluster} + u) P
\end{align}
where $U_{cluster}$ indicates that in $H_{lat}$,  Hubbard interactions are present only on the cluster sites, as in CDMFT, 
while the one-particle operator $u$ is used for lattice sites outside the cluster, i.e. there are no interactions in the bath.
%% (One can also define the  quantum impurity Hamiltonian as $H'= P (h + U) P$ ; this approximation generally performs similarly to the one above)~\cite{Knizia2013_JCTC9-1428-1432,Scuseria_Arxiv}.
Solving for the ground-state of $H'$ is a many-body problem with twice the
degrees of freedom as the impurity cluster.
%% ; for cluster sizes in this work (2-12) we use
 %% exact diagonalization, or quasi-exact density matrix renormalization group~\cite{dmrg}.
The resulting  quantum impurity wavefunction yields expectation values, such as energies and correlation functions that approximate those of the original lattice problem.
$u$ is adjusted to minimize the difference between the single-particle density matrix $\langle a^\dag_i a_j\rangle$ of   $H'$ and the ground-state of the model lattice Hamiltonian  $h+u$ projected to the impurity plus bath space, analogous to the self-consistent update of the self-energy in DMFT.
%% \begin{align}
%% \min_{u} ||\langle a^\dag_i a_j\rangle_{H'} - \langle a^\dag_i a_j\rangle_{h+u}||_{i,j \in \text{imp + bath}}
%% \end{align}
%% The resulting self-consistency is analogous to the self-consistent update of the self-energy in 
%% DMFT, and replicates the cluster DMET correlations to the bulk lattice. 

\begin{figure}[!t]
\includegraphics[width=8.8cm]{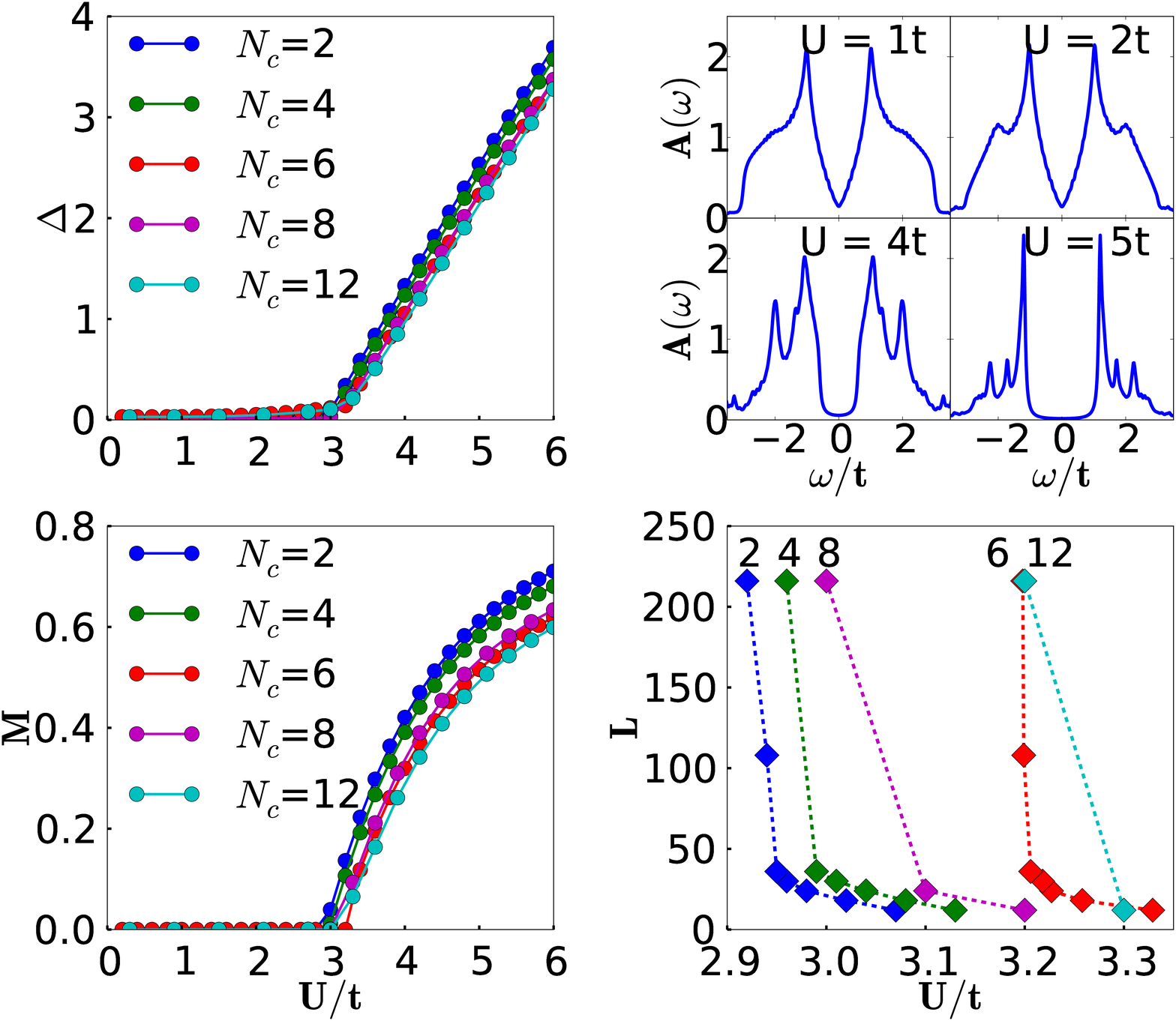}
\caption{(Color online) DMET calculations allowing spontaneous AFM order. Top Left: Single-particle gap against $U$ for cluster sizes $N_c=2-12$. As $U$ increases, a gap spontaneously opens at $U_{\rm AFM1}$, just after $U=3$. Top Right: Local density of states when $N_{c}=6$, calculated from spectral DMET~\cite{Booth2013_ArXive-prints1309-2320}, showing
the Dirac cone in the SM phase.
Bottom Left: Staggered magnetization against $U$ for $N_c=2-12$. The staggered magnetization appears at the same point as the
opening of the gap.
Bottom Right:  $U_{\rm AFM1}$ as a function of lattice size $L=12-216$ ($288-93312$ sites)
and  cluster size ($N_c=2-12$).}
\label{fig:uhf}
\end{figure}

To obtain spectra in DMET a modified procedure is used, where the ground-state bath space $\{ |\beta^0\rangle\}$ is augmented to reproduce dynamical properties.
%% by the states obtained by the Schmidt decomposition of the response vector
%%  of interest. 
 For example, to compute the local single-particle Green function, we consider the response vector of the model lattice wavefunction, 
\begin{equation}
|\Psi^{(1)}(\omega)\rangle=\frac{1}{\omega + \mu - h + u +  i0}a^{(\dagger)}_{\alpha}|\Psi^{(0)} \rangle
\end{equation}
and obtain a set of {\it additional} bath states from the Schmidt decomposition of $|\Psi^{(1)}(\omega)\rangle$, leading to a total bath space $\{ |\beta^{0}\rangle \oplus |\beta^1(\omega)\rangle\}$. The cluster Green function is determined by solving, at each frequency, the  response problem for $H_{lat}$ projected into the 
impurity plus dynamical bath space. Although the bath space is finite, it changes with frequency,  thus the finite impurity model produces a {\it continuous} spectrum along the real axis without artificial broadening.

We now turn to  applying DMET to the honeycomb Hubbard model.
%% Of primary interest are the phases at different $U$ and the order of the transitions
%% between phases. 
To identify different phases we monitor several quantities.
The SM phase is characterized by a vanishing single-particle gap and a Dirac cone 
at low energies in the single-particle density of states $A(\omega)$.
%% = -\frac{1}{\pi} \Im [ \sum_{\sigma} \langle \Psi^{(0)} | a_{\sigma} [\omega + \mu - (H-E_0) + i\eta]^{-1} a^{\dagger}_{\sigma} | \Psi^{(0)} \rangle  +  \langle \Psi^{(0)} | a^{\dagger}_{\sigma} [\omega + \mu + (H-E_0) + i\eta]^{-1} a_{\sigma} | \Psi^{(0)} \rangle ]$.
The AFM phase is characterized by non-vanishing  staggered
magnetization,  $M= \frac{1}{N} \sum_{ij}(-1)^{i+j}
(n_{i\uparrow}-n_{j\downarrow})$, a non-zero
 single-particle gap (which grows with 
increasing $U$), and a vanishing spin-gap. 
The proposed intermediate gapped phase is identified
 by a non-vanishing single-particle and spin-gap without long-range AFM order. We further check for spin-liquid character
from the correlation functions.
 Meng {\em et al.}~\cite{Meng2010_Nature464-847--851} 
further suggested that the intermediate phase can be identified from the gradient 
of $dT/dU$, where $T$ is the kinetic energy.
Not all the above quantities have been of equal focus in earlier studies;  AFQMC studies  report energies and gaps but not
 spectral functions, while  GFC calculations report spectral quantities 
but not energies. Here, using DMET, we study energies as well as  gaps and 
spectral functions.

%% report both 
%% directly compute ground-state quantities such as energies, and spectral
%% functions are hard to obtain, although estimates of the spin-gap can be obtained by computing in separate singlet and triplet sectors, while gaps can be inferred indirectly
%% from the correlation functions. Since AFQMC calculations directly produce ground-state energies, these and their derivatives (such as $dT/dU$) are further
%% used to characterize the phases. By contrast, the quantum cluster calculations are targetted at the single-particle density of states. This together with magnetization is the main tool
%% used to identify the phases and energies and spin-gaps are not reported. The gapped spin-liquid phase is identified by the opening of the single-particle gap without the development
%% of magnetic order. Indeed, Hassan and Senechal have conjectured that the spin-gap phase is in fact nothing other than the paramagnetic insulating phase typically observed in quantum cluster calculations at large $U$; certainly this phase has a non-vanishing single-particle and spin gap without long-range magnetic order. The paramagnetic insulating phase is not a ``ground-state'' phase
%% in the quantum cluster calculations, but can be observed if one does not allow magnetization to develop. This is commonly studied in cluster dynamical mean-field theory on the
%% square Hubbard lattice at half-dilling, where the paramagnetic insulating phase appears around $U=6$ although if magnetization is allowed to develop, the AFM phase appears at $U=0$.

To study the effect of cluster size and shape, we perform calculations
with $N_c=2-12$ cluster sites; these are shown in Fig.~\ref{fig:cluster}. For $N_c=2-6$ ($4-12$ sites with the bath) we use an exact diagonalization solver; for $N_c=8-12$ ($16-24$ sites with bath) we use a density matrix renormalization group (DMRG) solver~\cite{Sharma2012_J.Chem.Phys.136-124121}, keeping up to 2000 states, sufficient for quasi-exactness. 
It is important to distinguish cluster size $N_c$ from the size of the lattice ($2L^2$) in which the cluster is embedded. 
To study finite-size scaling effects of the underlying lattice, we use embedding lattices with $L \times L$ supercells ($2L^2$ sites), 
with $L=12-216$ (up to over 90,000 sites). The smaller $L$ calculations allow direct comparison to finite AFQMC calculations: for comparison, Meng {\it et al.} 
used up to  $L=18$; Sorella {\it et al.} used up to $L=36$.

%% Here we use our cluster  DMET formalism to compute both ground-state energies and correlation functions as well as dynamical spectral functions, allowing
%% direct comparison both to the prior AFQMC calculations as well as the quantum cluster calculations.
%% To begin, thus we map out the phase diagram using cluster DMET calculations, with cluster sizes $N_c=2,4,6,8$. The geometry of the clusters is shown in Fig. \ref{fig:cluster_shape}. To confirm that our results are converged to the thermodynamic limit, we consider the underlying model lattice to consist of two-site primitive unit cell and $L \times L$ supercells, resulting in the total number of sites (and electrons) in any system given by $2L^2$. We use up to $L=144$ or over 40,000 sites, to eliminate finite size lattice error. Note that
%% the computational time for the ground-state calculations was essentially the time for a set of exact diagonalization calculation on $2N_c$ sites and ranged from [less than a second] for $N_c=2$
%% to [a few hours ] for $N_c=8$. The dynamic spectra computed with $N_c=6$ took a few hours to compute.
%% [Need these plots; I'm guessing what is going on here]

We start by scanning the phase diagram as a function of $U$, {\it allowing 
spontaneous antiferromagnetism to develop}. The single-particle gap, magnetization, and
density of states for $N_c=2-12$ are shown in Fig.~\ref{fig:uhf}. 
We see at small $U$  the single particle gap vanishes and the
density of states displays a low-energy Dirac cone, clearly 
demonstrating that the system is in the SM phase.
As we increase $U$ beyond a critical $U_{\rm AFM1}$ an AFM solution
to the DMET equations appears and a gap opens
in the spectral function (Fig.~\ref{fig:uhf}). Note that $U_{\rm AFM1}$ is the earliest point at which the AFM solution can be found but this does not indicate
a thermodynamic phase transition at this point, which requires a more detailed examination of the energies,   discussed below. At $U_{\rm AFM1}$,
the single-particle gap and magnetization vanish simultaneously. 
Fitting $M=|U-U_{\rm AFM1}|^\beta$ to the $N_c=6$ data gives a critical exponent $\beta=0.72$,  compared 
with $0.80\pm 0.04$ from the AFQMC calculations of Sorella {\it et al.}, and $\beta=1$ from mean-field. 
$U_{\rm AFM1}$ shows significant cluster and lattice size dependence, as seen in Fig.~\ref{fig:uhf};
for $N_c=6$,  $L=12-216$, $U_{\rm AFM1}$ decreases from $3.329$ to $3.198$, while at $L=216$, for $N_c=2-12$, $U_{\rm AFM1}$ increases from $2.92$ to $3.2$.
Our $U_{\rm AFM1}=3.2$ for $N_c=12$, $L=216$ is somewhat lower than the recent AFQMC result $U_{\rm AFM1}=3.869$ of Sorella {\it et al.}~\cite{Sorella2012_SciRep2-992}.
The lattice size dependence of $U_{\rm AFM1}$ gives an estimate of the finite size error in  Sorella~{\it et al.}'s $L=36$ calculations; from $L=36-216$, 
$U_{\rm AFM1}$ decreases by $\sim  0.01$.

%% Our best estimate for $U_{\rm AFM1}$ appears well converged for $N_c=12$, $L=216$.
%% %% Further, the nature of the transition appears slightly different in the different
%% %% clusters, as seen from the different curvature of the magnetization near $U_{\rm AFM1}$ for $N_c=4,6$ in Fig.~\ref{gap}.
%% Carrying out finite size scaling with respect to $N_c$ and $L$ yields a bulk transition point of $U_{\rm AFM1}=[X]$, which compares fairly well
%% with the 

\begin{figure}[!t]
\includegraphics[width=8.8cm]{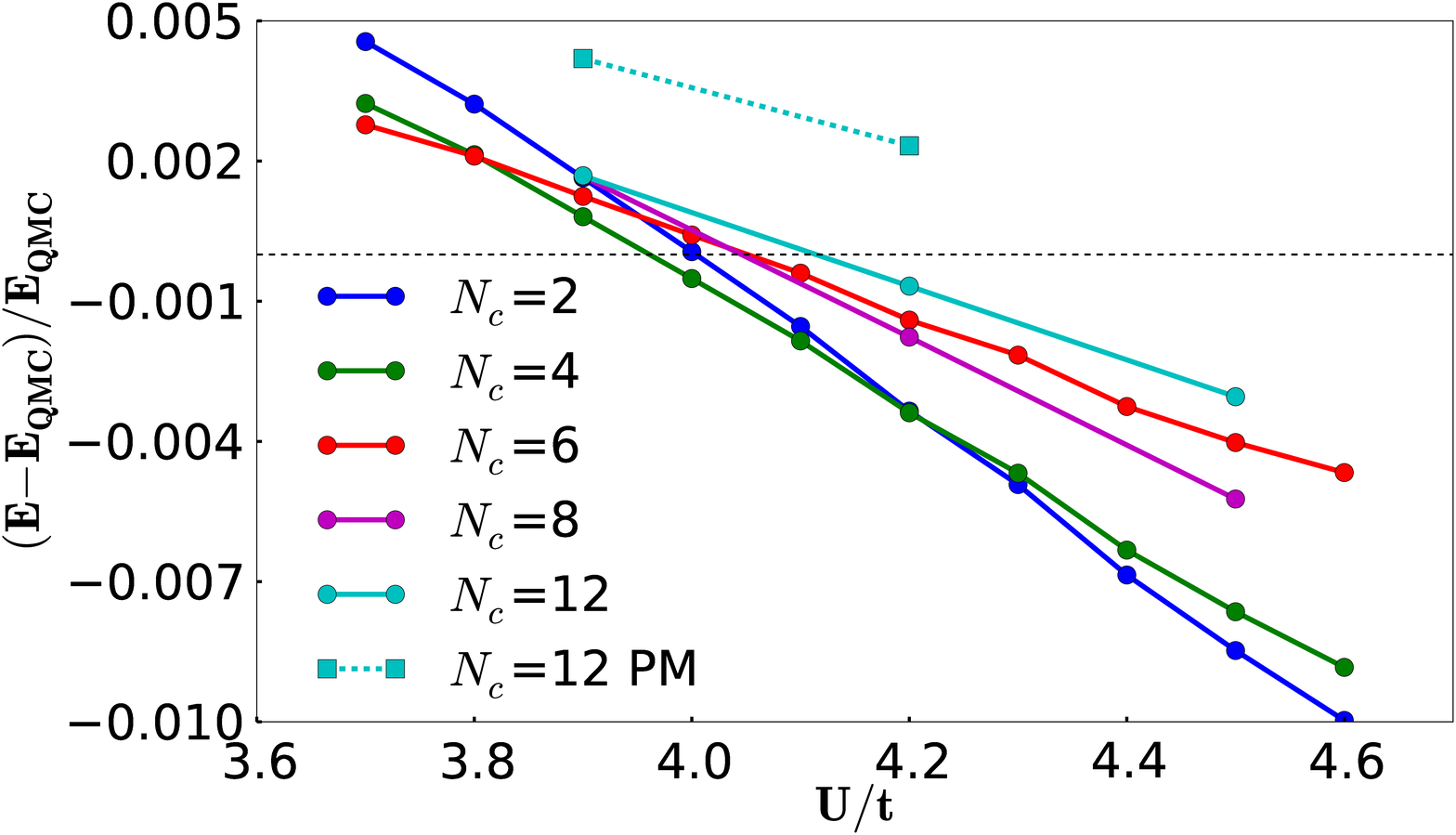}
\includegraphics[width=8.8cm]{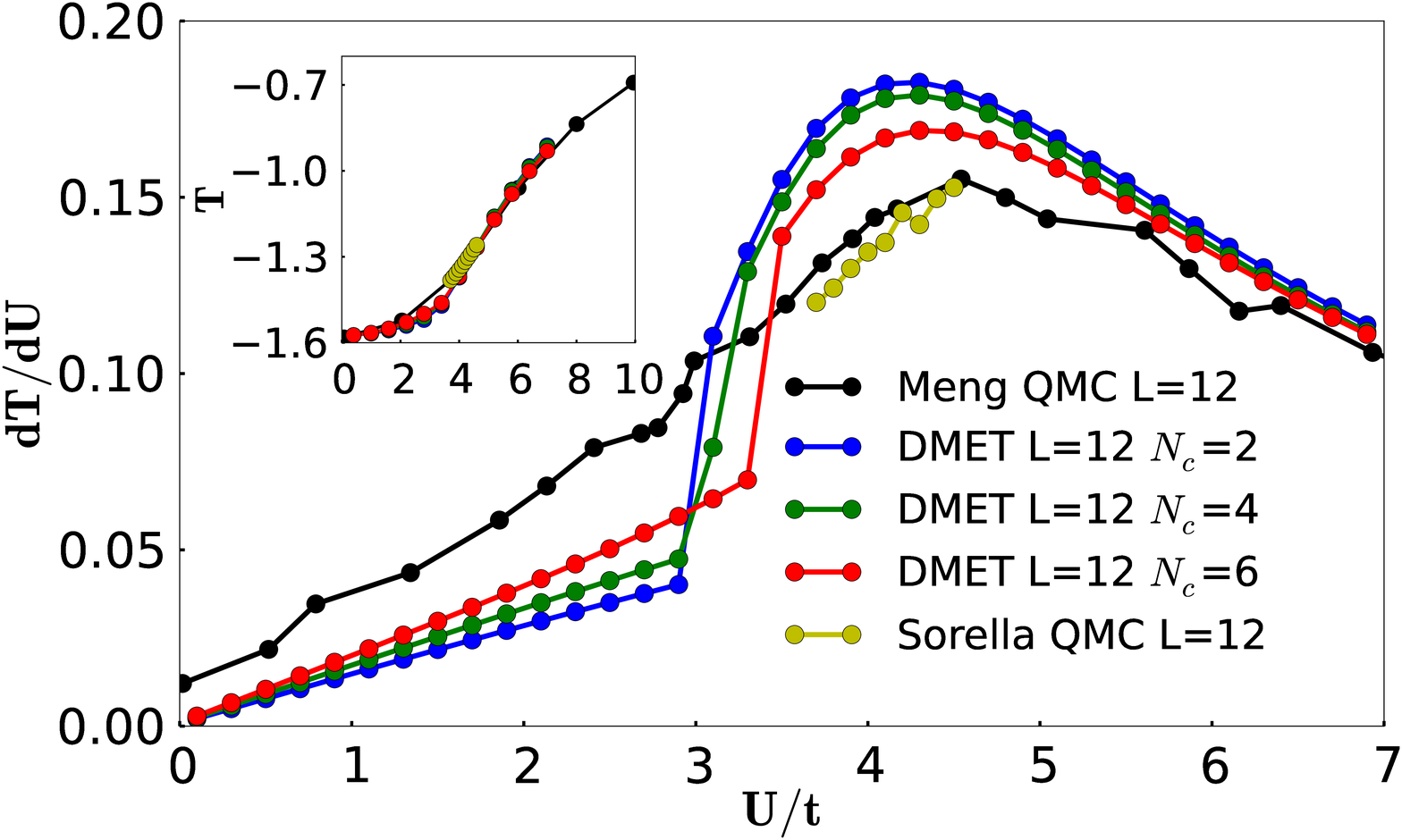}
\caption{(Color online) Top:  DMET energy versus Sorella et al.'s finite lattice AFQMC energy at intermediate $U$ for $L=12$ (288 sites) as
a function of cluster size $N_c=2-12$. This
is the most challenging regime for DMET, but the energy comparison is very favourable. PM denotes paramagnetic solution.
Bottom: Derivative of  kinetic energy as a function of $U/t$. Overlaid are results of Meng {\em et al.} and Sorella {\it et al.} for $L=12$~\cite{Meng2010_Nature464-847--851}.
Inset: Kinetic energy density as a function of $U$, showing good agreement between the DMET and numerically exact AFQMC~\cite{Meng2010_Nature464-847--851}.
DMET qualitatively reproduces  two changes in curvature (near $U=3.5$ and $U=4.3$) cited as evidence of an intermediate phase by Meng {\it et al.}, although
no intermediate phase is observed.}
\label{fig:energy}
\end{figure}

\begin{figure}[!t]
\includegraphics[width=8.8cm]{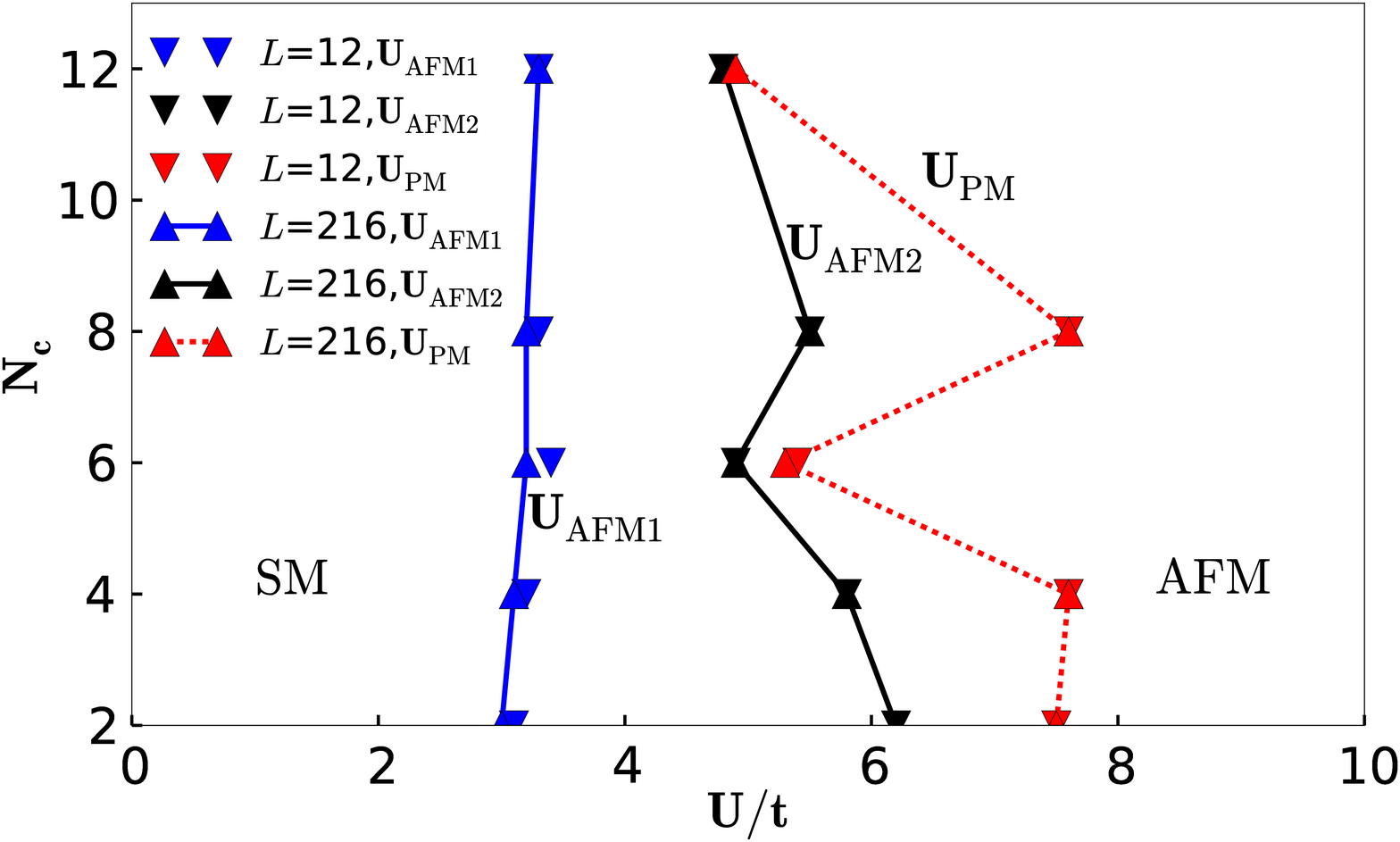}
\includegraphics[width=8.8cm]{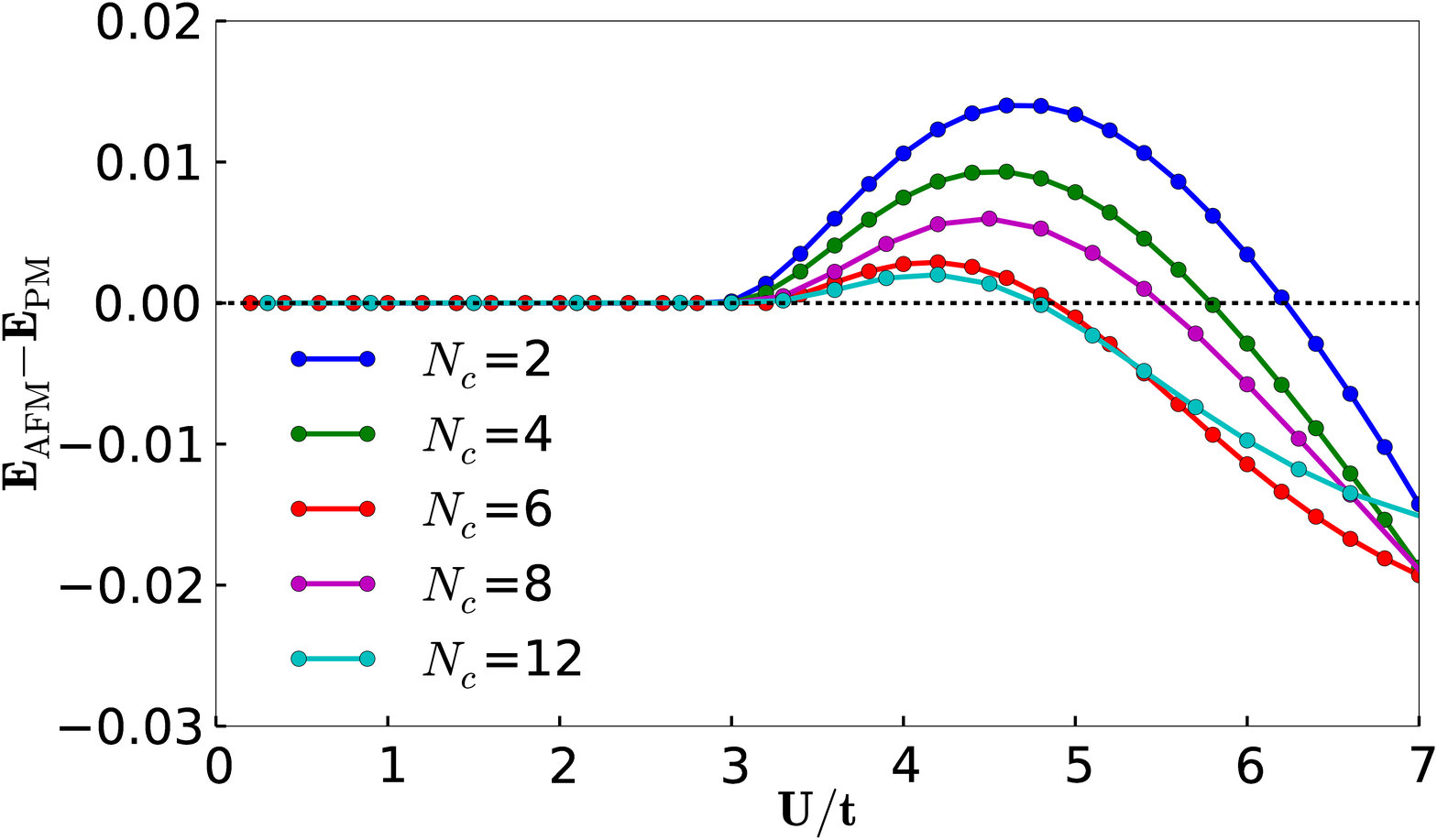}
\caption{(Color online) Top: Phase transition points as a function 
of cluster size 2-12, and two lattice sizes $L=12, 216$. $U_{\rm AFM1}$ corresponds
to opening of the  gap. $U_{\rm AFM2}$ corresponds
to the thermodynamic transition to the AFM phase. $[U_{\rm AFM1},U_{\rm AFM2}]$ is 
a coexistence region for the AFM phase and SM phase, although
this region appears to vanish with increasing cluster size. $U_{\rm PM}$ is the transition to the PMI if AFM order is not allowed to develop. The transition is first order for $N_c=2,4,8$ and second
order for $N_c=6,12$. Note that for $N_c=6,12$,  $U_{\rm PM}$ is very close to $U_{\rm AFM2}$ indicating that the PMI
is very competitive with the AFM phase for these cluster shapes.
Bottom: Ground state energy difference of PM and AFM solutions. The positive region
is the coexistence region.}
\label{fig:pd}
\end{figure}

A quantitative check of the accuracy of our calculations is to directly compare the energies for a given finite size lattice with the exact AFQMC energies for the same lattice size. This comparison for
total energies is shown in Fig. \ref{fig:energy}. We have compared over only a limited range of $U$
as this is where we had access to the QMC data~\cite{Sorella2012_SciRep2-992}. However, theoretically and numerically
it is known that the error in the DMET energies vanishes exactly at weak and strong coupling; the 
largest errors are near phase transitions~\cite{Knizia2012_PRL109-186404}, the region tested in Fig.~\ref{fig:energy}. 
Even in this challenging region, the cluster DMET total energies appear extremely good, as expected from earlier benchmarks~\cite{Knizia2012_PRL109-186404,Knizia2013_JCTC9-1428-1432,Scuseria_Arxiv};  with $N_c=12$ the energies are within 
$0.2$\% of the exact results. The more sensitive kinetic-energy derivative $dT/dU$ reported by Meng {\it et al.} (and obtained from Sorella {\it et al.})
is also shown in Fig. \ref{fig:pd}. Here we find that although Meng {\it et al.}
argued that the two changes in curvature in $dT/dU$ near $U=3.5,4.3$ indicate an intermediate phase, we also observe two changes
in curvature in our cluster DMET curve, at similar places, with just a SM and AFM phase, showing this is not a good diagnostic for an intermediate phase.

\begin{figure}[!t]
\includegraphics[width=8.8cm]{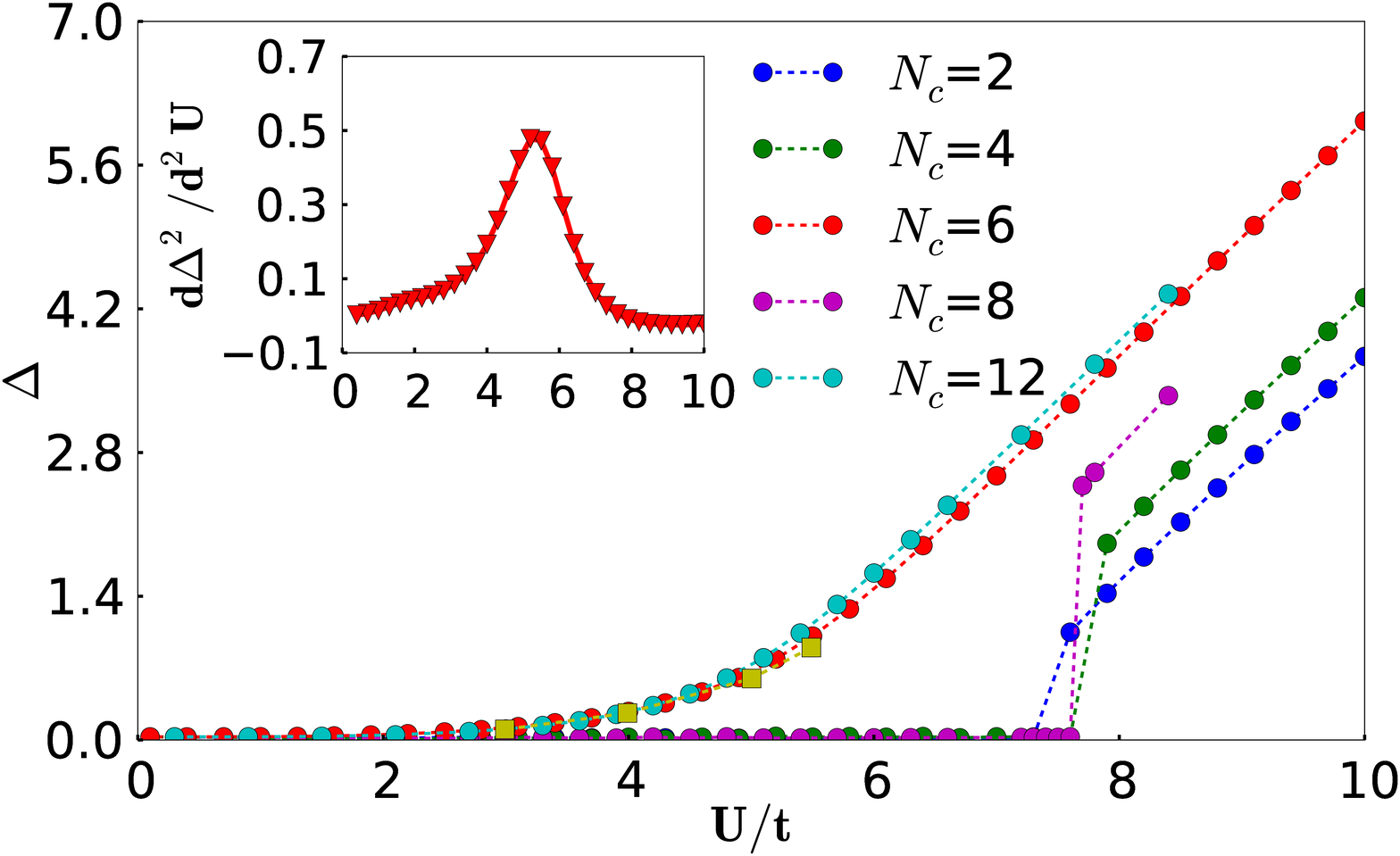}
\caption{(Color online) Single particle gap of paramagnetic solution with cluster sizes for $L=216$. For $N_{c}=2,4,8$ the transition is first order;
for $N_{c}=6,12$ the transition is continuous. Inset: second derivative of the gap for $N_c=6,12$; we identify
$U_{\rm PM}$ from the peak. Yellow squares: 6-site CDMFT calculations from Ref.~\cite{PhysRevB.87.205127}.}
\label{fig:rhf}
\end{figure}

We now consider the intermediate $U$ region and look for evidence of possible {\it metastable} intermediate phases.
To do so, we restrict our DMET calculations to paramagnetic phases and increase $U$. Interestingly, as $U$ is increased, 
we observe two kinds of paramagnetic transitions depending on cluster shape: for $N_c=2,4,8$ there is a first-order transition to a gapped phase, while
for $N_c=6,12$, there is a second-order transition: the gap opens continuously. 
In either case, we find a paramagnetic insulating phase (PMI) at large $U$.
%% This indicates a second order transition at a critical $U_{pm}$ to a (Fig.~\ref{gap}). The second-order nature of 
%% the transition means that it is {\it very} sensitive to the cluster size {\it and} the underlying lattice size. Indeed, as $N_c=2-12$, $U_{pm}=[X]-[Y]$,
%% while for fixed $N_c=6$, as $L=6\to 144$, $U_{pm}=[X]-[Y]$.
This finding of a PMI phase is not unexpected; preceding 6-site GFC calculations 
have identified a similar phase, typically identified as a Mott insulator. 
We can compare the  $N_c=6$ DMET PMI gaps directly to
earlier 6-site paramagnetic GFC calculations (see~\cite{PhysRevB.87.205127} for a summary). The agreement at large $U$ is good, but DMET
does not produce the  spurious gap seen in some GFC calculations at small $U$ that has led to  erroneous conclusions that there is no SM phase~\cite{PhysRevLett.110.096402}.
Note further, that we find the paramagnetic transition $U_{\rm PM}$ is very sensitive to cluster size, as seen in Fig.~\ref{fig:rhf}.

Hassan and Senechal conjectured that the
purported spin-liquid phase in the honeycomb Hubbard model {\it is} in fact the PMI phase in a quantum cluster calculation~\cite{PhysRevLett.110.096402}. We
now examine this conjecture. 
To identify whether the PMI phase is a true intermediate phase, we must check its stability in the presence of antiferromagnetism.
In the square lattice, antiferromagnetism order appears at infinitesimal $U$ before the PMI phase is reached.
The honeycomb Hubbard model differs in this regard because AFM order develops at finite $U_{\rm AFM1}$, which could in principle be larger than $U_{\rm PM}$.
%% the appearance of an  intermediate phase depends on the relative stability of the AFM phase and the PM phases at intermediate $U$. 
In Fig.~\ref{fig:pd} we show the energy of the AFM phase relative to the SM/PMI phases as a function of cluster and lattice size.
Here, we find that for our cluster sizes the transition between SM and AFM does {\it not} in fact occur at $U_{\rm AFM1}$, rather there is a first-order 
coexistence region between $U_{\rm AFM1}$, $U_{\rm AFM2}$, where $U_{\rm AFM2}$ is strongly cluster size (and to a lesser degree lattice size) dependent (Fig.~\ref{fig:pd}). 
As the cluster size increases, the first-order coexistence region decreases, and it appears that in the
limit of infinite cluster size the true AFM transition is second-order (Fig.~\ref{fig:pd}).
Further, in all instances we have studied, $U_{\rm AFM2}<U_{\rm PMI}$ at the same cluster size. This means that there is in fact {\it no} stable intermediate 
paramagnetic phase. However,
for the special cluster sizes $N_c=6,12$, the PMI phase is particularly competitive, and for these clusters, $U_{\rm AFM2}$ is very close to $U_{\rm PM}$.
Since locating $U_{\rm PM}$ precisely is itself difficult due to its continuous nature, a small uncertainty in $U_{\rm PM}$ could then yield $U_{\rm PM}<U_{\rm AFM2}$,
and a conclusion that there is a true paramagnetic intermediate phase.
In addition, the very small difference between $U_{\rm AFM2}$ and $U_{\rm PM}$ provides
a basis to explain the conflicting observations, in both AFQMC and in GFC calculations, of an intermediate phase near the antiferromagnetic transition: 
small changes in  calculation parameters can selectively stabilize the PMI phase. Our arguments therefore support the interpretation
of Hassan {\it et al.} that observations of the ``intermediate'' phase can be identified with the PMI state in quantum cluster calculations.

What is the nature of the metastable paramagnetic state for $N_c=6,12$? Although this ``intermediate phase'' is gapped without long-range magnetic order, 
this does not mean that it is a spin-liquid; another obvious candidate would be some kind of valence-bond crystal. The particular stabilization of the 
paramagnetic insulator for the hexagonal based clusters $N_c=6,12$, suggests that it is associated with a hexagonal cluster (valence-bond crystal) order. 
The annotations in Fig. \ref{fig:cluster} show the spin-spin correlation functions $\langle S^z_{i} S^z_{j}\rangle$. Although these correlation
functions in cluster DMET (as in CDMFT) are not guaranteed to preserve translational invariance, the pattern of translational invariance breaking
can be revealing of an underlying order. Indeed, the spin-correlation functions for $N_c=12$ confirm that a hexagonal cluster order develops in this
PMI phase (note there is no symmetry breaking in the corresponding SM phase for $N_c=12$, and further that the single $N_c=6$ cluster shows no evidence of dimerization). Intriguingly, hexagonal cluster order has been implicated
as a real instability of graphene under strain~\cite{Marianetti2010_PRL105-245502}. The (uncompetitive) PMI phase for $N_c=2,4,8$  develops a simpler dimer order.
%and one-particle density matrix elements $\langle a^\dag_i a_j\rangle$  on the cluster bonds. Indeed we find the clusters develop 
%% valence-bond like order in the intermediate phase. However, as the cluster size
%% increases, the valence-bond order begins to decrease. This  that the intermediate phase (if it manifests) is a weak valence-bond crystal
%% or a spin-liquid.

In conclusion, we have carried out cluster DMET calculations, as a function of cluster and lattice size, to elucidate
the phase-diagram of the half-filled Hubbard honeycomb model. Our detailed calculations find
that at intermediate couplings, there is a metastable paramagnetic insulating phase, that is very competitive with the antiferromagnetic phase. 
This insulating phase displays an associated hexagonal cluster order.
The closeness of the two phases at intermediate couplings
 means that small changes in calculational details can significantly affect their relative stability, and this 
explains the large number of conflicting results regarding the intermediate phase. 
It further seems likely that the intermediate paramagnetic phase can be stabilized by introducing modest frustration. Finally our work demonstrates the potential of the DMET methodology, which allows computation of both  energies and spectral functions at the thermodynamic limit, without incurring additional numerical artifacts.

\begin{acknowledgements}{We gratefully acknowledge S. Sorella and Y. Otsuka for providing the
energy data from their QMC calculations. This work was funded by the US Department of Energy, primarily through DE-SC0008624 and DE-SC0010530. Additional funding was provided through the DOE-CMCSN program DE-SC0006613.}
\end{acknowledgements}

\end{document}